\documentclass[%
 aip, 
 amsmath,amssymb, 
 reprint 
]{revtex4-1}
\usepackage{graphicx,dcolumn,bm}
\usepackage{xcolor} \usepackage[utf8]{inputenc} \usepackage[T1]{fontenc} \usepackage{mathptmx}
\usepackage{etoolbox} 

   \newcommand{\vomega}{\mbox{\boldmath $\omega$}}  
 \newcommand{\be}{\begin{equation}}   \newcommand{\ee}{\end{equation}}

\makeatletter
\def\@email#1#2{%
\endgroup
\patchcmd{\titleblock@produce}
{\frontmatter@RRAPformat}
{\frontmatter@RRAPformat{\produce@RRAP{*#1\href{mailto:#2}{#2}}}\frontmatter@RRAPformat}  {}{}
}%
\makeatother
    
\begin{document} 
\title{The role of field correlations on turbulent dissipation}
\author{Annick Pouquet} 
 \affiliation{$^1$Laboratory for Atmospheric \& Space Physics, University of Colorado Boulder, 80309 USA}  
\begin{abstract}
Nonlinear phenomena and turbulence are central to our understanding and modeling the dynamics of  fluids and plasmas, and yet they still resist analytical resolutions in many instances. However, progress has been made  recently, displaying a richness of phenomena which was  somewhat unexpected a few years back, such as the double constant-flux cascades of a same invariant to both the large and to the small scales, or the presence of non-Gaussian wings in the large-scale fields, for fluids and plasmas. 
Here, I will concentrate on the direct measurement of the magnitude of dissipation and an evaluation of intermittency in a turbulent plasma using exact laws stemming from invariance principles
{and involving cross-correlation tensors with both  the velocity and the magnetic fields.}
 I will  illustrate these points through scaling laws, together with data analysis from existing experiments,  observations and numerical simulations. Finally, I will also briefly explore the possible implications for validity and use of several modeling strategies. 
{\it To appear, Plasma Physics \& Controlled Fusion, 2023.}
 \end{abstract}  \maketitle


\section{Introduction}

 Observations in space at MHD scales, and later at proton and electron scales into the realm of plasmas, have yielded a wealth of data (see {\it e.g.} for recent analyses 
\cite{stawarz_16short,wang_19,matteini_20short} and references therein). Features of MHD turbulence 
using the Voyager spacecraft gave the scaling properties of magnetic energy and magnetic helicity $H_M \equiv \left<{\bf a} \cdot {\bf b}\right>$, with 
 ${\bf b}= \nabla \times {\bf a}$ the magnetic field \cite{matthaeus_82}.
 Later missions, like Cluster  \cite{matthaeus_05}, 
 led to discoveries such as field-aligned filaments  \cite{alexandrova_04short}, the rolling-up of Kelvin-Helmholtz (KH) vortices  \cite{hasegawa_04short}, or the  anisotropy of turbulent spectra in the magnetosheath \cite{sahraoui_06p}. Other space missions including THEMIS \cite{eriksson_09short}, MMS \cite{burch_16short,ergun_16short}, or more recently the Parker Solar Probe   \cite{bandyopadhyay_20short} have also yielded a wealth of results, making the assessment of turbulence in these plasmas more detailed as time progressed \cite{matthaeus_21}. 
For example, a recent MMS observation of reconnection in the magnetosheath \cite{stawarz_22short} indicates that electron-only reconnection is possible (see also \cite{daughton_11short,sahraoui_13short,zank_14,phan_18short}). It leads to very thin current sheets, as well as super-Alfvénic electron jets, in particular for short magnetic correlation length (based on the  correlation function of the magnetic field fluctuations).

Turbulence occurs for large kinetic and magnetic Reynolds numbers 
$Re=U_0L_0/\nu, \ R_M=U_0L_0/\eta$, with $U_0, L_0$ large-scale 
velocity and length scales, and $\nu,\eta$ the kinematic viscosity and magnetic resistivity.
Plasmas cover a wide range of spatio-temporal scales for which numerous dynamical models have been derived. MHD flows are more complex than incompressible fully developed turbulence (FDT)  because of the presence of waves which can alter the spectral energy distribution in the weak turbulence regime \cite{galtier_00}, as well as in the strongly nonlinear case \cite{iroshnikov_63,kraichnan_65} (IK; see also \cite{mininni_07b} for a recent discussion on the relevance of anisotropy in the absence of imposed uniform field).

The  Taylor scale is defined as  
\be
\lambda_V=[\left<u^2\right>/\left<\omega^2\right>]^{1/2}, \ee 
with 
$\vomega=\nabla \times {\bf u}$ the vorticity. It is located deep in the inertial range, between $L_0$ and the dissipative scale, taken {\it e.g.} as $\ell_d=[\epsilon_V/\nu^3]^{-1/4}$ for a Kolmogorov-1941 (or K41) kinetic energy spectrum $E_V(k)\sim \epsilon_V^{2/3}k^{-5/3}$, with $\epsilon_V\equiv D_tE_V$ the rate of dissipation of $E_V$. A magnetic Taylor scale $\lambda_B=[\left<b^2\right>/\left<j^2\right>]^{1/2}$ can be defined as well. Both are now measured in the solar wind \cite{matthaeus_05,bandyopadhyay_20bshort} and in laboratory plasmas \cite{cartagena_22}, allowing for a deeper understanding of scale interactions in these  flows. 

One can also define coherent structures which are of a size comparable to the integral scale, but which are also non-local in scale in the sense that they can be as thin as dissipation allows ({\it e.g.} 
$\sim \ell_d$ for fluids). Thus, their very existence implies that nonlinear interactions are  non-local as well. For example, in a fluid computation on a grid of $2048^3$ points forced with the Taylor-Green vortex, and at a Taylor Reynolds number of $R_\lambda=U\lambda/\nu \approx 1300$, the scale separation between forcing and dissipation allowed for the formation of several distinct wavenumber ranges. A slow decrease of the Kolmogorov constant was observed with increasing $R_\lambda$, together with an increase of both the skewness and kurtosis of velocity increments, and with more intermittency (that is, more departure from a self-similar scaling) than in the classical She-Leveque model \cite{mininni_08} (see \cite{alexakis_07b} for the MHD case). Note also that the intermittency in the far dissipative range is found to involve large-scale structures and differs from that in the inertial range, a result that is obtained both through direct numerical simulations (DNS) and with a classical model of turbulence \cite{chen_93}.
  
To tackle these problems, two main research tools are DNS on the one hand, and analytical weak (wave) turbulence (WT) on the other hand. Their development will be mostly ignored here, except for the brief following reminders. Although there were several precursors, in particular  in two space dimensions (2D) or in the kinematic (fixed velocity) case, the first accurate non-linear 3D incompressible MHD  numerical computation of MHD turbulence (at least, in front of the fence) was performed starting in 1975 \cite{pouquet_78p}. The computations followed the path opened by S.A. Orszag and G.S. Patterson  \cite{patterson_71,orszag_72b} using  pseudo-spectral methods for fluid turbulence. Many improvements were accomplished over the years, one of which is adaptive mesh refinement on a grid  \cite{friedel_97} or in a spectral context \cite{rosenberg_06}, the degree of refinement  depending on which norm ({\it e.g.}, ${\cal L}_2$ {\it vs} ${\cal L}_\infty$) is sought to be modeled accurately  \cite{rosenberg_07,ng_08}.

Turbulence being nonlinear, the formulation of the problem in terms of  moments 
is incomplete, with one more unknown than there are equations. Closures have thus to be devised, which are justified in the presence of a small parameter. Diagrammatic techniques were introduced to study fully developed fluid turbulence, for example the Direct Interaction Approximation (DIA) \cite{kraichnan_59} (see also \cite{krommes_15} for plasmas), or for weak fluid or plasma  turbulence \cite{vedenov_61,benney_62,hasselmann_62-1,zakharov_65} (see \cite{zakharov_92,nazarenko_11b} for introductions). In WT,  the closure is exact but non-uniform in scale. Indeed, the small parameter of the problem is the ratio of the wave period to the eddy turn-over time $\tau_{wave}/\tau_{NL}$ with $\tau_{NL}=L_0/U_0$. It is small for fast waves, but in general it does not remain small as scales change. For MHD, this parameter can be taken as the ratio of Alfvén times built respectively on the large uniform imposed magnetic field and the fluctuating field.

The WT formulation for MHD in 3D was derived in \cite{galtier_00} (for unbalanced plasmas, see {\it e.g.} \cite{passot_19}). Nonlinear diffusion equations can be deduced from the full set of the closed WT formalism, including in the case of strong cross helicity  defined as $H_C=\left< {\bf u} \cdot {\bf b} \right>$ \cite{galtier_10}. When the WT formulation breaks down, not all is lost and closures have been proposed such as the DIA mentioned above, and the (simpler) Eddy Damped Quasi-Normal Model (EDQNM), as developed for MHD in the non-helical \cite{kraichnan_67n} and in the helical \cite{pouquet_76j} cases. The relevance of such closures, even today, is that they are more easily numerically integrated at high Reynolds number thanks to the possibility of an exponential discretization in wavenumber. One might also note that nonlinearities are not strong everywhere even in the absence of waves: indeed, turbulent flows produce structures that are quasi force-free (parallel current and induction), Beltrami (parallel vorticity and velocity), or Alfvénic (parallel velocity and magnetic field). This may be important for plasmas since field-aligned currents can be unstable \cite{stawarz_15j}. This Beltramization of fluids and plasmas nonlinearly creates structures weakening the nonlinearities, making them survive longer than expected, that is beyond a few eddy turn-over times. 

In the remainder of the paper, I shall simply review  the use of so-called exact laws in determining the dissipation (\S \ref{S:EXACT}) and intermittency properties (\S \ref{S:INTER}) of turbulence, and I shall then give some rapid conclusions and perspectives.

\section{Measuring dissipation 
through exact laws} \label{S:EXACT}  
In a highly turbulent flow, the energy dissipation rate 
$\epsilon_V\equiv \nu \left<{\bf u} \cdot \Delta {\bf u}\right>$  (with averages taken over the volume, say) is of order the dimensional estimate $\epsilon_D=U_0^3/L_0$.
This has been demonstrated with numerical and experimental data for fluids \cite{sreenivasan_98}, and in MHD when examining the total energy dissipation $\epsilon_T=\epsilon_V+\epsilon_M, \epsilon_M=\eta \left<{\bf b} \cdot \Delta {\bf b}\right>$ with $\epsilon_M$  the magnetic energy dissipation \cite{mininni_09}. 
Furthermore, the conservation properties of ideal invariants also led to the derivation of  exact laws, as first shown by Kolmogorov \cite{K41a} for the kinetic energy in 3D fluids, initially under several hypotheses (isotropy, incompressibility, homogeneity, stationarity and finite dissipation at high $Re$). This law relates the third-order two-point  structure function of the longitudinal component  of the velocity $\left<\delta u_L^3(r)\right> , $ (with 
$\delta {\bf u}({\bf r}) = {\bf u}({\bf x}+\bf{r}) - {\bf u}({\bf x})$, $u_L={\bf u} \cdot {\bf r}/r$),
to $\epsilon_V$ and the distance $r=|r|$ between the two points. In fact, two different laws can be written, where the second formulation includes all components of the velocity \cite{antonia_97}: 
\be  \left<\delta u_L^3(r)\right> = -\frac{4}{5} \epsilon_V r, \ 
S_3^V\equiv \left<\delta u_L \Sigma_i \delta u_i^2 \right> = -\frac{4}{3} \epsilon_V r \ , \ee 
 neglecting forcing and dissipation. The flux can be positive or negative, depending on the direction of transfer to small or large scales, or to both as in MHD with a strong uniform field, or in rotating stratified flows \cite{marino_15p,pouquet_13p,alexakis_18}. Similar coupled laws have been derived (for recent reviews, see \cite{coburn_15,banerjee_17,marino_23r}), for kinetic helicity as well as for MHD \cite{politano_98p,politano_98g}, Hall MHD and electron MHD \cite{galtier_08} for the total energy and cross-helicity invariants, with further extensions to  magnetic helicity and generalized helicity in Hall MHD, and to the compressible case  \cite{ferrand_21}.

 It can be shown that some of the hypotheses can be discarded (such as that of isotropy or of compressibility), in particular using a more compact vectorial flux formulation \cite{banerjee_17}. The resulting relationships involve vectorial forces (Lamb vector ${\bf u} \times \vomega$, Lorentz force and Ohms law ${\bf u} \times {\bf b}$), forces which may be more difficult to measure in space plasmas, and which are, moreover, weakened by coherent structures as stated before. Note that considering such invariants is essential to our understanding of the nonlinear dynamics insofar as they place strong physical and scaling constraints on the correlation functions of the flow; they play central roles \cite{frisch_83}, in particular in reconnection events ({\it e.g.} \cite{pouquet_88,hosking_21}).
 
In the case of 3D incompressible MHD, one begins with writing the generalized von K\`arm\`an-Howarth equation in terms of $[{\bf u}, {\bf b}]$  correlation functions \cite{chandrasekhar_51,politano_98p}. One then derives exact scaling laws for the joint third-order structure functions of the velocity and magnetic field; this step involves {\it a priori} the kinematics of tensors that include both vectors and pseudo-vectors, ${\bf u}, {\bf b}$, and that are non-symmetric in their indices. The two coupled laws for MHD  involve mixed triple correlations; they read \cite{politano_98p,politano_98g}:
\begin{eqnarray} 
M_3^T \equiv \left<\delta u_L \Sigma_i [\delta u_i^2 + \delta b_i^2] \right> 
- 2 \left<\delta b_L \Sigma_i \delta u_i\delta b_i\right> &=& -4\epsilon_T r/3 , 
\label{exact-mhd1} \\
M_3^C \equiv - \left<\delta b_L \Sigma_i [\delta u_i^2 + \delta b_i^2] \right> 
+ 2 \left<\delta u_L \Sigma_i \delta u_i\delta b_i \right> &=&-4\epsilon_C r/3 , \label{exact-mhd2} 
\end{eqnarray} 
where $\epsilon_{T,C}$ are the dissipation rates of the total energy and the velocity-magnetic field  cross-correlation. These laws  can  be written in a more symmetric form in terms of the Elsasser variables, ${\bf z}^\pm ={\bf u} \pm {\bf b}$ (see \cite{politano_98p,politano_98g}). 

Examining the equations above, there are obviously three regimes: one dominated by the velocity in which one recovers the three-component exact law for fluids \cite{antonia_97}, one dominated by the magnetic field where one obtains again a Kolmogorov-like scaling, with presumably a -5/3 law for an (isotropic) magnetic energy spectrum, and an intermediate, mixed regime ($b \sim u$) in which the correlations between ${\bf u}$ and ${\bf b}$ are dynamically  important. One should note that (a) because of intermittency, the scaling with $r$ of structure functions of higher order cannot be deduced simply from the third-oder law. MHD is known to be more intermittent than fluids so that, even in the magnetically-dominated regime, one does not expect (and one does not find) similar intermittency departure from self-similarity than with the velocity \cite{politano_98e,sorriso_04short,abramenko_10}.
And (b): this analysis is independent of the fact that magnetic helicity is also conserved in ideal MHD.
The first confirmation of these laws was obtained in 2005 \cite{macbride_05}, and in 2007 using Ulysses data \cite{sorriso_07short} (see also \cite{marino_08} for a measure of the dissipation itself), studies which were pursued further to include the high-latitude Solar Wind in the presence of strong velocity shear \cite{marino_12}.
 
Extension to the compressible case has led to estimates for the dissipation rates of both the energy and the cross-helicity in the Earth magnetotail \cite{hadid_18}; measurable differences are observed compared to the incompressible case and this may have consequences for reconnection events. Finally, also note an interesting discussion in \cite{yousef_07} whereby the omitted viscous and resistive terms in the above laws may in fact play a role even at high Reynolds number, in particular when the magnetic Prandtl number differs from unity, this role being linked to the structuring of the magnetic field into stripes.
 These pioneering results were  followed by many observational, numerical and theoretical developments. A recent  example in the Solar Wind and the magnetosheath concerns the effect of adding new terms to the above laws corresponding to the contributions stemming {\it e,g,} from the Hall current \cite{bandyopadhyay_20short}. 
In fact, when looking at approximations of Hall-MHD, for example involving finite Larmor radius effects, one could search for similar exact laws and their effect on the stability properties of solitary waves (see {\it e.g.} \cite{bello_19}).

Another remarkable and perhaps unexpected feature of these  observations is that the energy flux is seen to change sign at some scale \cite{sorriso_07short}, corresponding to a change of direction of the energy cascade. A confirmation of the results concerning the change of sign of the flux at a given scale \cite{sorriso_07short}, in the exact law for total energy in MHD, is to be found in \cite{hernandez_21short} using recent Parker Solar Probe data in the inner heliosphere.
This may hint at an inverse magnetic helicity cascade, with the magnetic energy having to grow as well at large scale because of a Schwarz inequality constraint, but this would need to be checked.

\section{Exact laws and measurements of intermittency} \label{S:INTER}
Beyond the mere pleasure of writing new algebra leading to new exact scaling laws, and of measuring with novel tools and increased accuracy the dissipation of turbulent flows in space plasmas, these exact laws may also be useful in assessing more precisely small-scale intermittency due to the presence of strong localized structures such as current sheets and filaments. Intermittency is analyzed classically through the non-Gaussian wings of the gradients of the velocity, magnetic field and temperature fields, examining the skewness and kurtosis of their Probability Distribution Functions (PDFs) as well as through the scaling of varying-order structure functions.

However, intermittency is known to be difficult to assess since it involves a scaling analysis of high-order structure functions, in particular for odd orders when sign changes occur. Assuming  a simple scaling, one can write  $\left<\delta f^n\right> \sim r^{\zeta^f_n}$, where the  $\zeta^f_n$ are  the 
$f$-field scaling exponents. Complete self-similarity would give $\zeta^f_n=a_p n$, with $a_p$ a coefficient depending on dimensional analysis ($a_p=1/3$ for K41, and $a_p=1/2$ for the isotropic IK total energy law \cite{iroshnikov_63,kraichnan_65}). Intermittency leads to a nonlinear variation of 
$\zeta^f_n$ with $n$, corresponding to non-Gaussian wings in the PDFs of $f$ \cite{grauer_94,politano_95a}. It is linked to  coherent structures that are strong and (somewhat) isolated, such as vortex tubes, current sheets and current filaments, as well as switch-backs, in the vicinity of which dissipation can become very high. For example, a recent study using the Parker Solar Probe shows that such structures account roughly for 19\% of measurements of intermittency in the solar wind, and that the small-scale eddies are directly associated with an increase in proton temperature viewed as a proxy for dissipation \cite{sioulas_22short}. There are numerous indications that $\zeta^f_n$ is not linear in the order $n$ of the structure functions, for Burgers turbulence \cite{gotoh_99}, for fluids \cite{ishihara_09}, and for space plasmas (see {\it e.g.} \cite{bruno_05} and references therein). 

A remark here seems to be in order, namely that the idea of extended self-similarity (or ESS, \cite{benzi_93}) has already been generalized to the analysis of solar-flare data. In ESS, the structure function in the exact law within the inertial range is used as a proxy for the distance r in plotting 
$\left<\delta u^n(r)\right>$, and similarly for MHD using for the distance r, eqations (\ref{exact-mhd1},\ref{exact-mhd2}) (see \cite{sorriso_02,mininni_09} for such a case on DNSs of $1024^2$ and $1536^3$ points respectively). Note that one has also used a one-field third-order structure function based on either the velocity field or the Elsasser fields instead of the exact laws written above for MHD 
\cite{mueller_00,uritsky_07,uritsky_17p}. 
Better scaling has been obtained using ESS because of the intrinsic correlations in the flow \cite{benzi_93,abramenko_10,uritsky_10,klimas_17}, and yet this simple idea may not have been fully exploited when analyzing space and numerical data, for either the energy, the magnetic helicity or the generalized Hall-MHD helicity (see \cite{kutsenko_18} and references therein for current helicity intermittency in solar active regions, focusing on the flatness). 

For example, the exact law in flux-form for the symmetrized two-point correlation function for magnetic helicity conservation takes the compact form 
\be M_3^{\eta_M}\equiv \left <\delta({\bf u} \times {\bf b}) \cdot \delta {\bf b} \right>  
- d_i \left<\delta({\bf j} \times {\bf b}) \cdot \delta {\bf b} \right> = \eta_M\equiv DH_M/Dt \ , \ee 
with $d_i$ the ion inertial length scale  \cite{banerjee_16}. Note that these authors also give the exact law for the generalized magnetic helicity (using ${\bf b} + d_i \vomega$), and that the assumption of isotropy is not necessary in these flux derivations. Using  expressions for the mixed MHD third-order correlators $M_3^{T,C}, M_3^{\eta_M}$ should however give  a more accurate representation of the possible variations of scaling exponents with governing parameters, for high-resolution DNS including in 2D, as well as for solar wind, magnetospheric and experimental data.

The advantage of taking the mixed correlation function as it appears in the exact laws is that it combines the two ${\bf u,b}$ or ${\bf z}^\pm$ variables, and thus it directly incorporates  the amount of correlation between the velocity and the magnetic field, a key player in the dynamics. These laws have already been observed in DNS of MHD flows in 2D and 3D \cite{politano_98e,mininni_07,malapaka_13}. Note that one of the two $M_3^{T,C}$ correlators might yield better results, depending on the amount of cross-correlation in the flow. Such an analysis might also give a better representation of the role of small (sub-ion and sub-electron) scales in the shaping of coherent structures. Of course, the conservation of other ideal invariants such as the squared magnetic potential $\left<A^2\right>$ in 2D, or the generalized helicity (a second helical invariant in Hall MHD), $H_G=H_M+2 d_iH_C + d_i^2 H_V$, with $H_V= \left<{\bf u} \cdot \vomega \right>$
the kinetic helicity, allows for similar direct measures of their own dissipation and intermittency, as well as their topology. Related issues will arise in the presence of ambipolar drift as it occurs in molecular clouds  \cite{momferratos_14}, and the corresponding exact laws can be found useful as well to determine more accurately the nature of intermittency in the interstellar medium, which is linked 
{to shear \cite{falgarone_90,falgarone_09,momferratos_14},}
 to shocks and to vorticity and current sheets \cite{vazquez_06}.
 
Several issues arise that will affect the amount of dissipation occurring in such flows, and the nature of intermittency leading to non-Gaussian PDFs. For example, a highly compressible one-dimensional (1D) gas does not tend to have Burgers-like statistics because of strong density and pressure fluctuations \cite{passot_98}. The model developed by these authors could be extended to MHD and Hall-MHD, possibly contrasting it with the 1D models already developed in the seventies by analogy to Burgers equation \cite{thomas_70}. One should also note the similarity between the exact law for total energy for Hall-MHD and the possible existence of a dissipation anomaly in that case for the local energy balance in the limit of infinite Reynolds number \cite{galtier_18}.

Models of turbulence such as the Lagrangian averaged (alpha) model or the EDQNM, also possess exact laws that can be analyzed, as done in \cite{briard_18} for the EDQNM in MHD in the presence of strong cross-helicity. For the MHD alpha model, a break in the scaling of the third-order law is visible at the $\alpha$-scale at which the modeling strongly comes into effect
 \cite{mininni_05p,graham_06}. Using this third-order law as $r$  might yield interesting results on intermittency in the case of the model, both for fluids and for MHD, contrasting them with DNS, 
 {and thus allowing for much higher equivalent Reynolds number simulations \cite{graham_11p}.}

As a final note, intermittency of the large scales (velocity, induction, temperature{, density}), as opposed to the small scales (vorticity, current, temperature or density gradients) has also been found in many contexts in the presence of waves \cite{paoletti_08,sardeshmukh_15,pouquet_20r,chau_21short}, as well as in the solar wind \cite{greco_09,marino_12} and in magnetic storms \cite{wang_22}. For example, in stratified turbulence, the vertical velocity can have high kurtosis in a narrow range of the governing weak turbulence parameter, {\it i.e.} the Froude number \cite{feraco_21short}, and it can make the dissipation even more localized than for fully developed turbulence \cite{marino_22p}. A non-Gaussian kurtosis for the magnetic field, which depends on scale, is also observed in the solar wind and it can be as high as several hundreds \cite{chasapis_17short}. Similarly, interactions between shear and zonal flows have been studied for a long time in the context of the so-called L-H transition in plasmas  \cite{diamond_94,itoh_06short,gurcan_15}, and here again the  role of vertical velocity shear is central in the large-scale intermittent dynamics \cite{smyth_19,pouquet_19p}.

\section{Conclusions and Perspectives}  \label{S:CONCLU} 

Many  topics have been omitted in this paper, such as particle acceleration or, as important, 
the dynamo, {\it i.e.} the generation of magnetic fields by turbulence, and the role of helicity. 
In the latter case, numerous laboratory and numerical experiments have been performed (see {\it e.g.} 
\cite{ponty_07p,nore_18,taylor_86,childress_95,brandenburg_05,tobias_21} for reviews). 
Small-scale dynamos  are possible even in the absence of helical forcing \cite{mininni_05a}, but in fact magnetic helicity plays a central role in the nonlinear phase of the dynamo since $H_M$ can be viewed as the motor of the large-scale growth of a magnetic field being entrained by the inverse cascade of $H_M$, and allowing as well for saturation \cite{pouquet_76j}. Both magnetic and cross-helicity have been shown recently in more detail to play a role in the dynamo problem and in reconnection \cite{guo_12}, and their effect on small-scale development should be incorporated in models  \cite{yokoi_13}.  A recent study of imbalanced flows in the specific case of local interactions of kinetic Alfvén waves from MHD to sub-ion scales, shows that the generalized cross-helicity $H_G$ can undergo a direct or an inverse cascade \cite{passot_19}. Moreover, these authors show that the forcing term included in the problem can influence the outcome of the small-scale dynamics (see also \cite{passot_18}). 

Another important topic mostly omitted in this review is that of  modeling of turbulent flows, a necessity without which we would not be where we are today. The parallel development of DNS and models in the early 70s was central to the unraveling of the small-scale properties of fluids and plasmas. One way to extend the inertial range of turbulence is to increase the order of the dissipation operator, using for example hyper-viscosity. This may be particularly useful for supersonic flows for which both spectral \cite{passot_87,passot_88c} and implicit numerical modeling methods have been developed, such as The Piecewise Parabolic Method \cite{colella_84,porter_94}.  

One can construct a hierarchy of models from quasi-linear to three-mode interactions to simplified large-scale flows as a zonal flow \cite{li_22}, and assess the contribution of each approach to the energy balance, as well as estimating characteristic triple-correlation timescales. Comparisons have been performed in MHD where one could, for example, quantify to some extent the contribution of kinetic or cross helicity to transport coefficients \cite{yokoi_13}, or the role of incorporating new decorrelation times in the eddy-damping process, such as the departure from an assumed kinetic-magnetic energy equipartition \cite{baerenzung_08b}. 

Today, DNS are paramount, and in view of the gigantic technical progress made, more complex physics can be directly tackled with computers. There is a palpable feeling that computers can do it all, that the modeling can be deduced in an automatized way, as opposed to the systematic, if algebraically cumbersome, techniques developed starting in quantum physics and moving to nonlinear fluid and MHD problems, through the use of Feynman diagrams and renormalisation methods in general as well as (somewhat {\it ad hoc}) closure models; all have been reviewed at length elsewhere. To take only a few recent examples, one can pinpoint the onset of the formation of strong current structures, from a fully kinetic formulation to that of MHD, as  for the KH instability, using wavelets \cite{le_18}. One can also  identify closure terms in a hierarchy of moment equations from numerical simulations by performing a nonlocal closure for the electron heat flux, learning from 3D kinetic simulation data in the case of reconnection in a so-called double-Harris sheet using a particle in cell code, the input being the large-scale fields and their gradients as well as density and pressure ratio \cite{laperre_22} (see  \cite{shukla_22} for another example).

Renewed recourse to closures such as the EDQNM or the nearest-neighbor shell models built on conservation properties (see {\it e.g.} \cite{alexakis_18} for a recent review), might be of use in examining high $Re$ plasma turbulence in the presence of waves for some of the problems mentioned here. Indeed, the persistence or not of dual bi-directional cascades at substantially higher Reynolds numbers would be a candidate for study, and what determines their relative strength, as well as the large-scale intermittency and its concomitant dissipation and scaling with parameters. For example, using the EDQNM, it is shown in \cite{briard_18} that the presence of a strong cross-helicity leads to the development of anisotropy and to changes in spectral behavior. These authors also assess exact scaling laws within the framework of the closure, and find excellent agreement.  

In conclusion, if some of the ideas evoked herein are classical, the data produced by new and soon-to-come spacecrafts and high Reynolds number numerical simulations together with modeling is {and will be} phenomenal in its increased resolution and data stream. {These tools} will continue to provide immense testing grounds for the study of intricacies of small-scale transport models in fusion and turbulent plasmas, and on the role waves such as (kinetic) Alfv\'en waves, play in modulating {the localization and intensity of dissipative structures,} the transport  {properties} and the scaling laws arising from intermittency, both at large and at small scales.

\vskip0.12truein

{\bf Acknowledgements}

{\sl I want to thank the many mentors, colleagues and students with whom I performed over many years the research referred to here, and more of course, principally in Nice and Boulder, principally using analytical closures and numerical simulations, and principally on MHD, later on exploring rotating stratified turbulence. I am also thankful to LASP, and in particular to Bob Ergun.} 

\bibliographystyle{apsrev4-1}      \bibliography{maastrich_revised.bib}     

 \end{document}